\documentclass[aps,prb,twocolumn,superscriptaddress,showpacs]{revtex4}

\usepackage{graphicx}% Include figure files
\usepackage{dcolumn}% Align table columns on decimal point
\usepackage{bm}% bold math

\begin{document}

\title{Low-temperature heat transport of the layered spin-dimer compound Ba$_3$Mn$_2$O$_8$}

\author{W. P. Ke}
\affiliation{Hefei National Laboratory for Physical Sciences at
Microscale, University of Science and Technology of China, Hefei,
Anhui 230026, P. R. China}

\author{X. M. Wang}
\affiliation{Hefei National Laboratory for Physical Sciences at
Microscale, University of Science and Technology of China, Hefei,
Anhui 230026, P. R. China}

\author{C. Fan}
\affiliation{Hefei National Laboratory for Physical Sciences at
Microscale, University of Science and Technology of China, Hefei,
Anhui 230026, P. R. China}

\author{Z. Y. Zhao}
\affiliation{Hefei National Laboratory for Physical Sciences at
Microscale, University of Science and Technology of China, Hefei,
Anhui 230026, P. R. China}

\author{X. G. Liu}
\affiliation{Hefei National Laboratory for Physical Sciences at
Microscale, University of Science and Technology of China, Hefei,
Anhui 230026, P. R. China}

\author{L. M. Chen}
\affiliation{Department of Physics, University of Science and
Technology of China, Hefei, Anhui 230026, P. R. China}

\author{Q. J. Li}
\affiliation{Hefei National Laboratory for Physical Sciences at
Microscale, University of Science and Technology of China, Hefei,
Anhui 230026, P. R. China}

\author{X. Zhao}
\affiliation{School of Physical Sciences, University of Science
and Technology of China, Hefei, Anhui 230026, People's Republic of
China}

\author{X. F. Sun}
\email{xfsun@ustc.edu.cn}
\affiliation{Hefei National Laboratory for Physical Sciences at
Microscale, University of Science and Technology of China, Hefei,
Anhui 230026, P. R. China}

\date{\today}

\begin{abstract}

We report the study on the low-temperature heat transport of
Ba$_3$Mn$_2$O$_8$ single crystal, a layered spin-dimer compound
exhibiting the magnetic-field-induced magnetic order or the magnon
Bose-Einstein condensation. The thermal conductivities ($\kappa$)
along both the $ab$ plane and the $c$ axis show nearly isotropic
dependence on magnetic field, that is, $\kappa$ is strongly
suppressed with increasing field, particularly at the critical
fields of magnetic phase transitions. These results indicate that
the magnetic excitations play a role of scattering phonons and the
scattering effect is enhanced when the magnetic field closes the
gap in the spin spectrum. In addition, the magnons in the BEC
state of this materials do not show notable ability of carrying
heat.

\end{abstract}

%66.70.-f Nonelectronic thermal conduction and heat-pulse propagation in solids
%75.50.-y Studies of specific magnetic materials
%75.47.-m Magnetotransport phenomena; materials for magnetotransport

\maketitle

\section{Introduction}

Low-dimensional or frustrated quantum magnets were revealed to
exhibit exotic ground states, magnetic excitations, and quantum
phase transitions (QPTs).\cite{Sachdev, Balents} For a particular
case of the spin-gapped antiferromagnets, the external magnetic
field can close the gap in the spectrum, which results in a QPT
between a low-field disordered paramagnetic phase and a high-field
long-range ordered one. An intriguing finding is that this ordered
phase can be approximately described as a Bose-Einstein
condensation (BEC) of magnons.\cite{BEC_Review} Many experimental
investigations on the critical properties of the BEC-related QPT
have been carried out, including the characterizations of the
magnon spectrum, the magnetization, the specific heat and the
thermal transport, etc.\cite{BEC_Review} In a recent study, the
heat transport properties of a magnon BEC material,
NiCl$_2$-$4$SC(NH$_2$)$_2$ (DTN), were found to display strong
anomalies at the QPTs and the heat conductivity of the BEC state
seemed to be much enhanced upon lowering temperatures (approaching
the absolute zero).\cite{Sun_DTN} This result shows an analogy
between the magnon BEC and the superfluid of $^4$He in the aspect
of the ability of transporting heat. However, one notable facet is
that magnons act as heat carriers only in the direction of the
spin chains of this compound, whereas they only scatter phonons in
the transverse direction.\cite{Sun_DTN} So the exchange anisotropy
may play the key role in the heat transport of magnetic
excitations. A later experimental work confirmed the main features
of transport properties of DTN, but an alternative picture based
on the mass renormalization and impurity scattering effects was
proposed to explain the thermal transport data.\cite{Kohama} In
any case, the QPTs associated with the magnon BEC are believed to
significantly affect the heat transport properties and the
low-energy magnetic excitations provide a substantial contribution
to the heat transport. To get the general principals of the heat
transport in the magnon BEC state, we need to carry out systematic
studies on more members of the magnon BEC materials.

Ba$_3$Mn$_2$O$_8$ (BMO) is an $S$ = 1 quantum spin-dimer system
exhibiting the BEC of spin degrees of freedom. It crystallizes in
the rhombohedral $R\bar{3}m$ structure with the pairs of Mn$^{5+}$
3$d^{2}$ ions ($S$ = 1) arranged vertically on the hexagonal
layers.\cite{Uchida1} The spins of Mn$^{5+}$ ions of each pair are
coupled antiferromagnetically to form spin dimers.\cite{Uchida2,
Tsujii, Xu} Neutron scattering results indicated an intradimer
exchange energy $J_0$ = 1.642 meV, an interdimer coupling between
Mn ions in the same plane $J_2 - J_3$ = 0.1136 meV, an interdimer
coupling between Mn ions in the adjacent planes $J_1 = -0.118$
meV, and the next nearest neighbor interdimer interactions between
bilayers (along the $c$ axis) $J_4 = -0.037$ meV.\cite{Stone1,
Stone2, Samulon1, Samulon2, Samulon3} The strong in-plane
spin-dimer interaction results in a spin-singlet ground state,
with a spin gap of 1.05 meV to the lowest triplet state and a
second larger gap to the quintuplet state. BMO displays a peculiar
phase diagram in magnetic field when the Zeeman effect lowers the
$S_Z$ = 1 triplet states and the $S_Z$ = 2 quintuplet
states.\cite{Samulon2} Corresponding to the closures of two spin
gaps with increasing magnetic field, there are two sequential
magnetically ordered states and four quantum critical fields:
$H_{c1}$ = 8.7 T, $H_{c2}$ = 26.5 T, $H_{c3}$ = 32.5 T, $H_{c4}$ =
47.9 T for $H \parallel c$.\cite{Uchida2, Samulon1, Samulon2} The
peculiarity is that the first ordered phase for $H_{c1} < H <
H_{c2}$ and the second one for $H_{c3} < H < H_{c4}$ are described
as the triplet condensation and the quintuplet condensation,
respectively.\cite{Samulon2} Furthermore, BMO was found to have a
weak single-ion uniaxial anisotropy, characterized by $|D|$ =
0.032 meV.\cite{Samulon3} The phase diagram for $H \perp c$ is a
bit more complicated, with two phases of the triplet
condensation.\cite{Samulon1, Samulon2, Samulon3} It was revealed
that both the single-ion anisotropy and the geometry frustration
play crucial roles in determining the phase
diagram.\cite{Samulon1}

In this work, we study the thermal conductivity ($\kappa$) of BMO
single crystal to probe the field-induced QPTs and the role of
magnetic excitations in the heat transport. It is found that both
$\kappa_{ab}$ and $\kappa_c$ are strongly suppressed in magnetic
fields, particularly at the field-induced QPTs, which demonstrates
that at very low temperatures the magnons mainly act as phonon
scatterers even in the long-range ordered state. The present
results indicate that magnon-BEC state may not necessarily exhibit
large thermal conductivity and the ability of magnetic excitations
transporting heat is determined by the magnetic structure and the
anisotropic spin exchange.

\section{Experiments}

High-quality single crystals of Ba$_3$Mn$_2$O$_8$ are grown by a
slowly cooling method using NaOH as a solvent.\cite{Samulon1} The
crystals have a hexagonal-platelet-like shape with large sizes up
to 5 $\times$ 5 $\times$ 2.5 mm$^3$. The largest surface is the
$ab$ plane and the thickness is along the $c$ axis, determined by
using the x-ray diffraction and the Laue photograph. The specific
heat is measured by the relaxation method in the temperature range
from 0.4 to 20 K using a commercial physical property measurement
system (PPMS, Quantum Design). Both the $ab$-plane and the
$c$-axis thermal conductivities ($\kappa_{ab}$ and $\kappa_c$) are
measured using the conventional steady-state technique.
Particularly, the data at low temperature and in high magnetic
field are taken by using a ``one heater, two thermometer" method
in a $^3$He refrigerator and a 14 T magnet.\cite{Sun_DTN,
Wang_HMO, Zhao_GFO} Note that we can only probe the QPTs at the
first critical field of BMO, due to the limitation of the
available field.

\section{Results and Discussion}

\begin{figure}
\includegraphics[clip,width=6.5cm]{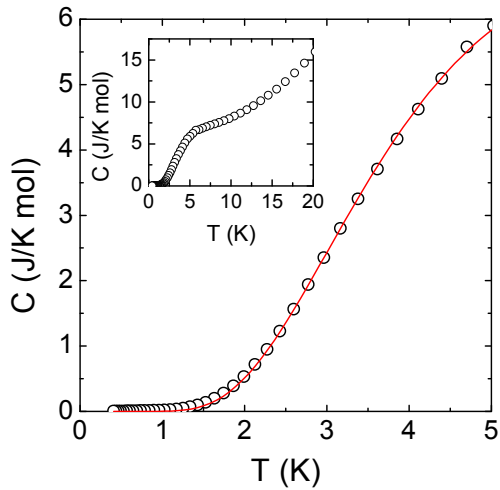}
\caption{(Color online) Low-temperature specific heat of
Ba$_3$Mn$_2$O$_8$ single crystal in zero field. The solid line
indicates the fit of data using formula (\ref{SH}). Inset: the
data in a broader temperature range from 0.4 to 20 K.}
\end{figure}

It is known that, in a magnetic insulator, heat can be carried by
phonons and magnetic excitations, and the interaction between them
usually induces scattering on heat carriers, which is a negative
effect on the heat transport. To analysis the heat transport
properties quantitatively, it is necessary to know the phononic
specific heat from the experiments. Figure 1 shows the low-$T$
specific heat of the Ba$_3$Mn$_2$O$_8$ single crystal. At $\sim$ 6
K, there appears a broad shoulder-like feature, which is
apparently of magnetic origin. Below 5 K, the specific heat
decreases quickly with temperature and it does not show a $T^3$
dependence of the phonon specific heat. The same result has been
reported in an earlier work.\cite{Tsujii} The low-$T$ specific
heat was found to be able to described by a formula for the gapped
magnetic excitations,\cite{Tsujii}
\begin{equation}
C=\tilde{n}R(\Delta/T)^{2}e^{-\Delta/T}, \label{SH}
\end{equation}
where $\tilde{n}$ is the number of excited states per spin dimer,
$R$ is the gas constant, and $\Delta$ is the energy gap of
magnetic excitations. The low-$T$ data can be fitted rather well
to Eq. (\ref{SH}) with parameters $\tilde{n}$ = 1.49 and $\Delta$
= 14.2 K, except for the very low-$T$ data ($<$ 1 K).\cite{Tsujii}
Note that the size of gap is quite consistent with those from
other measurements.\cite{Uchida2, Stone2} However, it is difficult
to separate the phononic specific heat because of the significant
contribution from the magnetic excitations. Actually, adding a
$T^3$-term to Eq. (\ref{SH}) cannot achieve perfect fitting to the
data, particulary for the subKelvin data.

\begin{figure}
\includegraphics[clip,width=8.5cm]{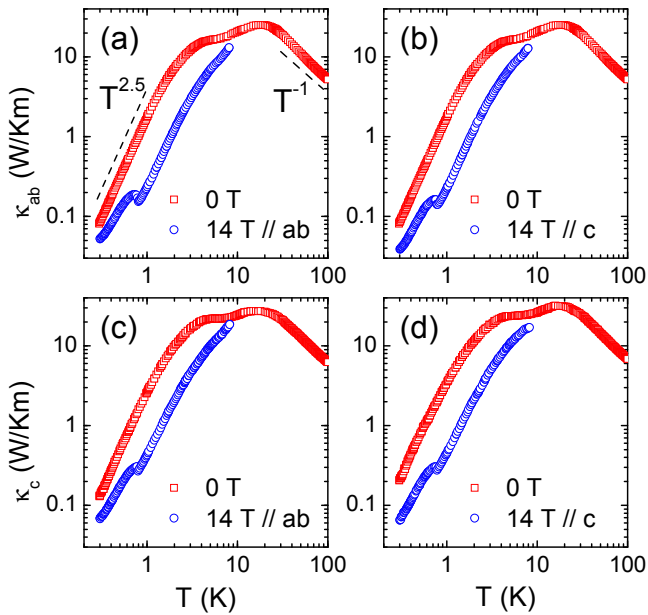}
\caption{(Color online) Temperature dependencies of the thermal
conductivities $\kappa_{ab}$ and $\kappa_c$ of Ba$_3$Mn$_2$O$_8$
single crystals in zero and $14$ T fields. The dashed lines
indicate the approximate $T^{2.5}$ and $T^{-1}$ dependencies of
the thermal conductivity at very low temperatures and at high
temperatures, respectively.}
\end{figure}

Figure 2 shows the temperature dependencies of $\kappa_{ab}$ and
$\kappa_c$ of BMO single crystals in zero and 14 T fields, which
are applied along either the $ab$ plane or the $c$ axis.
Apparently, the zero-field thermal conductivities show rather weak
anisotropy at high temperatures. Each $\kappa(T)$ curve exhibits a
phonon peak at 18 K with the peak values nearly the same for
$\kappa_{ab}$ and $\kappa_c$. The $T^{-1}$ dependence of
$\kappa(T)$ at high temperatures is the characteristic of phonon
heat transport dominated by the phonon-phonon Umklapp
scattering.\cite{Berman} At sub-Kelvin temperatures, both
$\kappa_{ab}(T)$ and $\kappa_c(T)$ show an approximate $T^{2.5}$
dependence, which indicates that the boundary scattering limit of
phonon thermal conductivity is approached or that the microscopic
scattering on phonons is negligible.\cite{Berman} Besides these
usual behaviors belonging to the phonon heat transport, there is a
``shoulderlike" feature in the zero-field $\kappa(T)$ at 6.5 K,
which is likely due to some kind of resonant scattering on
phonons.\cite{Berman} Applying 14 T magnetic field leads to a
strong suppression of low-$T$ thermal conductivity and the
disappearance of the ``shoulderlike" feature. Therefore, the
phonon resonant scattering in zero field must be of magnetic
origin. Similar result has been found in another spin-gapped
material, DTN.\cite{Sun_DTN} On the other hand, there is no
evidence for the magnon heat transport in BMO, which is reasonable
at very low temperatures considering the negligible magnon
excitations due to the finite spin gap. In high magnetic field,
the strong suppression of phonon heat transport in a broad
temperature range is apparently due to the enhancement of magnetic
scattering, which can also be seen from the following
magnetic-field dependencies of $\kappa$. Another peculiar feature
of 14 T $\kappa(T)$ data is the small jumps of conductivity at low
temperatures, which are related to the field-induced
antiferromagnetic (AF) transitions. Note that even with these
increases of $\kappa$ at the phase transitions, the lower-$T$
thermal conductivities are still much smaller than the zero-field
values, indicating that the phonons are still strongly scattered
by magnetic excitations in the field-induced AF state.

\begin{figure}
\includegraphics[width=8.5cm]{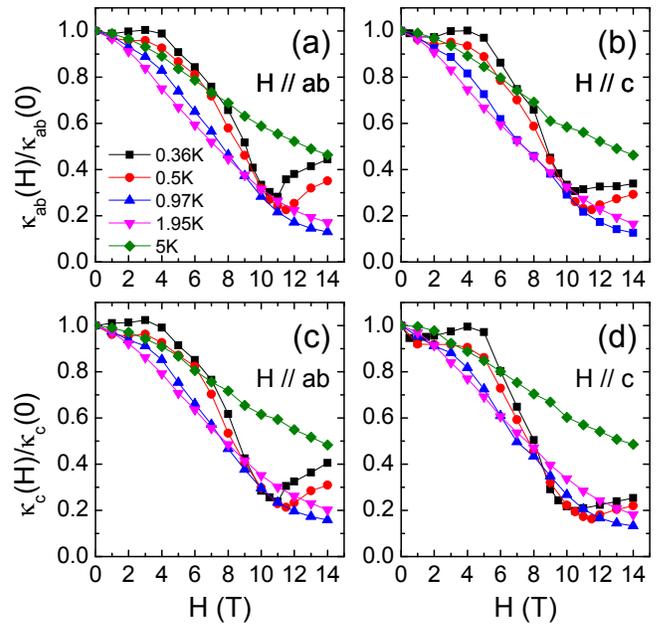}
\caption{(Color online) Magnetic-field dependencies of thermal
conductivity of Ba$_3$Mn$_2$O$_8$ single crystals at low
temperatures.}
\end{figure}

Figure 3 shows the detailed magnetic field dependencies of
$\kappa_{c}$ and $\kappa_{ab}$ for both $H \parallel ab$ and $H
\parallel c$. It is clear that they are nearly isotropic on the field
direction. The overall behavior of $\kappa(H)$ is that the
magnetic field strongly suppresses thermal conductivity, but at
very low temperatures, {\it i.e.}, at 0.36 and 0.5 K, $\kappa$
shows upturn above some critical fields. Since the zero-field
thermal conductivity of BMO is purely phononic, the field-induced
suppression of $\kappa$ is clearly a result of scattering effect
on phonons by some magnetic excitations. In this regard, the
phonon scattering by paramagnetic moments, which are commonly
existing as impurities or spin vacancies, is one of the
unavoidable effects.\cite{Berman, Sun_GBCO, Sun_PLCO} However, the
field dependencies of $\kappa$ shown in Fig. 3 are rather
different from the standard one of paramagnetic moments
scattering.\cite{Berman, Sun_GBCO} The main origin for the
$\kappa(H)$ behaviors is therefore the phonon scattering by magnon
excitations. The low-field quantum disordered phase has a finite
spin gap, which can be weakened by the Zeeman effect. At a fixed
temperature, with increasing the magnetic field, the spin gap
decreases and the number of low-energy magnons increases quickly,
which can strongly scatter phonons and leads to a significant
suppression of $\kappa$.

A notable feature of $\kappa(H)$ isotherms is the upturn of
$\kappa$ in high magnetic field and at very low temperatures. This
increase of $\kappa$ is obviously related to the field-induced AF
order since the transition fields are found to be coincided with
the phase boundary of high-field ordered phase, {\it i.e.}, the
magnon BEC state. There are two possible reasons for the increase
of $\kappa$ in the BEC state. First, the phonon scattering by
magnons is weakened when the magnon BEC occurs. Second, the
magnons in the BEC state have a positive effect on $\kappa$ by
transporting the heat directly. If the magnons transport the
energy, due to the anisotropy of the exchange and the
low-dimensionality of the spin structure, the field dependence of
$\kappa$ should exhibit obvious anisotropy on the direction of
heat current with the same field.\cite{Brenig, Hess, Sologubenko}
However, it can be seen from Fig. 3 that the upturn behaviors
actually show a weak anisotropy on the direction of magnetic field
rather than on that of the heat current. This means that the
upturn of thermal conductivity in the BEC state is a result of the
weakening of magnon-phonon scattering. In other words, although
the magnon scattering on phonons are generally strong at high
fields, it is more significant at the critical fields because of
the strong spin fluctuations.

Although the above discussions are quite clear in catching the
main physics of the observed heat transport properties, it is
useful to try a more quantitative analysis on the experimental
data. First, since the phonons are the only one type of heat
carriers, the zero-field $\kappa(T)$ data are likely to be fitted
by a classical Debye model of phonon thermal
conductivity\cite{Berman,Sologubenko2}
\begin{equation}
\kappa_{ph}=\frac{k_B}{2\pi^2v}(\frac{k_B} {\hbar})^3T^3
\mbox{\resizebox{1.5\width}{1.5\height}{$\int$}}_0^{\Theta_D / T}
\frac{x^4e^x}{(e^x-1)^2}\tau(\omega,T)dx, \label{kappa}
\end{equation}
in which $\omega$ is the phonon frequency, $x = \hbar \omega/k_BT$
is dimensionless, and $1/\tau(\omega,T)$ is the phonon relaxation
rate. The relaxation time is determined by
\begin{equation}
\tau^{-1} = \upsilon/L + A\omega^4 + BT\omega^3\exp(-\Theta_D/bT)
+ \tau^{-1}_{res} , \label{tau}
\end{equation}
which represent the phonon scattering by the grain boundary, the
point defects, the phonon-phonon Umklapp scattering, and the
resonant scattering, respectively.\cite{Berman} The average sound
velocity $v$ can be calculated from the Debye temperature using
the formula $\Theta_D = v(\hbar/k_B)(6\pi^2n)^{1/3}$, where $n$ is
the number density of atoms. For BMO, however, the Debye
temperature cannot be simply obtained from the phonon
specific-heat data, as the data in Fig. 1 indicate. For an
estimation, we refer the specific-heat data of
Ba$_3$V$_2$O$_8$,\cite{Samulon4} a nonmagnetic material having the
same crystal structure as BMO. The phonon specific-heat
coefficient $\beta = 6.06 \times 10^{-4}$ J/K$^4$mol can be easily
obtained from a $T^3$-fitting to the low-$T$ data.\cite{Samulon4}
With this $\beta$ value, the Debye temperature and the sound
velocity of BMO are calculated to be 346 K and 2900 m/s,
respectively. The parameter $L$ describing the boundary scattering
is the the averaged sample width of the samples. The other
parameters $A$, $B$ and $b$ are free ones.

One difficulty of getting a precise calculation using Eq.
(\ref{kappa}) is the description of $\tau^{-1}_{res}$. In BMO, the
magnetic excitations can strongly scatter phonons. Considering the
energy dispersion of magnons, it is very difficult to get a
transparent formula for $\tau^{-1}_{res}$.\cite{} For a simple
analysis, one can neglect the magnon dispersion and assume that
the resonant scattering of phonons occurs via a singlet-triplet
excitations of dimerized states. Thus, the resonant scattering
rate can be expressed as
\begin{equation}
\tau^{-1}_{res} = C\frac{\omega^4}{(\omega^2_0-\omega^2)^2}F(T),
\label{tau_res}
\end{equation}
where $C$ is a free parameter while $F(T)$ describes the
difference of thermal populations of the excited triplet and the
ground singlet states (the contribution of a higher-level
excitations of quintuplet states is negligibly small). It is known
that\cite{Sologubenko2, Stone2}
\begin{equation}
F(T) = 1 - \frac{1-exp(-\Delta/T)}{1+3exp(-\Delta/T)}, \label{F}
\end{equation}
with $\Delta = \hbar\omega_0$ the energy gap of magnetic spectrum,
which is about 14.2 K from the specific-heat data.

\begin{figure}
\includegraphics[width=6.0cm]{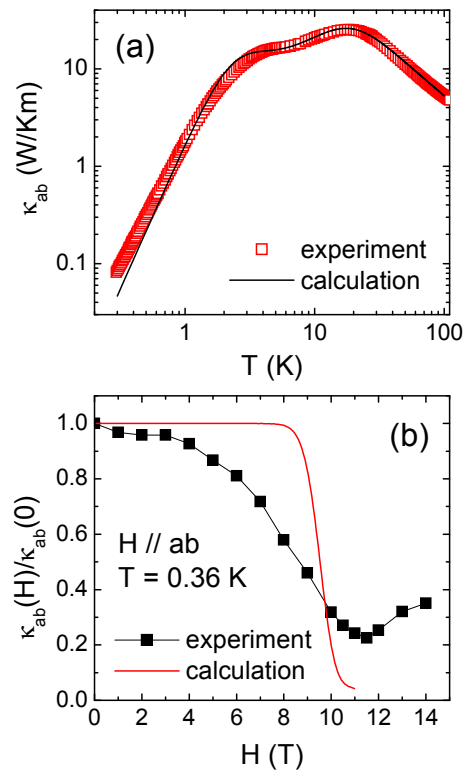}
\caption{(Color online) Comparison of the $\kappa_{ab}(T)$ data
(a) and the $\kappa_{ab}(H)$ data (b) with the calculation results
using the Debye model.}
\end{figure}

The zero-field thermal conductivity data are able to be fitted by
using the above formula very well. As an example, the zero-field
data from Fig. 2(a) and the fitting curve are displayed in Fig.
4(a). The best fitting parameters are $A$ = 1.3 $\times$
10$^{-42}$ s$^3$, $B$ = 3.0 $\times$ 10$^{-30}$ K$^{-1}$s$^2$, $b$
= 6.0, and $C$ = 1.6 $\times$ 10$^8$ s$^{-1}$.

Second, the magnetic-field dependence of $\kappa$ can also be
calculated using the Debye model. The effect of magnetic field is
to induce the Zeeman splitting of the degenerate triplet states
and a linear decrease of the energy gap with field, that is,
\begin{equation}
\Delta(H) = \Delta - g\mu_BH/k_B. \label{E}
\end{equation}
The Land\'{e} factor was known to be $\sim
2.00$.\cite{Stone2,Samulon1} A typical comparison between the
$\kappa(H)$ data and the calculations is shown in Fig. 4(b).
Although there is a qualitative agreement between the experimental
data and the calculation, the quantitative difference is quite
large. In particular, at low fields the experimental data show
much stronger suppression than the calculated result. It is
expectable from the classical Boltzmann distribution that the
magnon excitations are negligibly weak at subKelvin temperatures
when the spin gap is larger than 10 K. The experimental data
suggest that the magnon excitations, which may be because of the
quantum fluctuations, are much stronger than the thermal
excitation produces. Another reason for the significant
discrepancy is likely related to the simple assumption of the
resonant phonon scattering by the two-level magnetic excitations.
Actually, the magnon dispersions of BMO are not very weak, which
makes the scattering between phonons and magnetic excitations
inelastic and much more complicated. A precise calculation based
on the magnon-phonon scattering calls for the details of both the
phonon spectra and the magnon dispersions.\cite{Upadhyaya, White}

It is useful to compare these results with those of DTN, in which
the magnetic field along the $c$ axis (the spin-chain direction)
can close the spin gap and drive the AF phase transition. In DTN,
the heat transport along the $c$ axis or perpendicular to it
present remarkably anisotropic behaviors with increasing field,
that is, $\kappa_c$ show sharp peaks at $H_{c1}$ while
$\kappa_{ab}$ show dips at the same position.\cite{Sun_DTN} Those
demonstrated that magnons act as the heat carriers along the $c$
axis and the phonon scatterers along the $ab$ plane,
respectively.\cite{Sun_DTN} Furthermore, the $\kappa_c$ of DTN in
the AF state (for $H_{c1} < H < H_{c2}$) keep increasing with
lowering the temperature, also pointing to a potentially large
heat transport in the magnon BEC phase.\cite{Sun_DTN, Kohama} The
present $\kappa(H)$ behaviors of BMO for both $\kappa_{ab}$ and
$\kappa_c$ are very similar to those of $\kappa_{ab}(H)$ in DTN,
which suggests that the magnetic scattering on phonons are
enhanced upon approaching the critical fields. In contrast to the
case of DTN, the magnetic excitations of BMO do not contribute to
transporting heat substantially along either the $ab$ plane or the
$c$ axis. Therefore, whether the magnon BEC state of the
spin-gapped compounds can have strong ability of transporting heat
is mainly dependent on such factors as the nature of the magnetic
structure and the exchange anisotropy etc. The field-induced AF
state or magnon BEC is not the sufficient condition for observable
magnon transport.

\begin{figure}
\includegraphics[width=8.5cm]{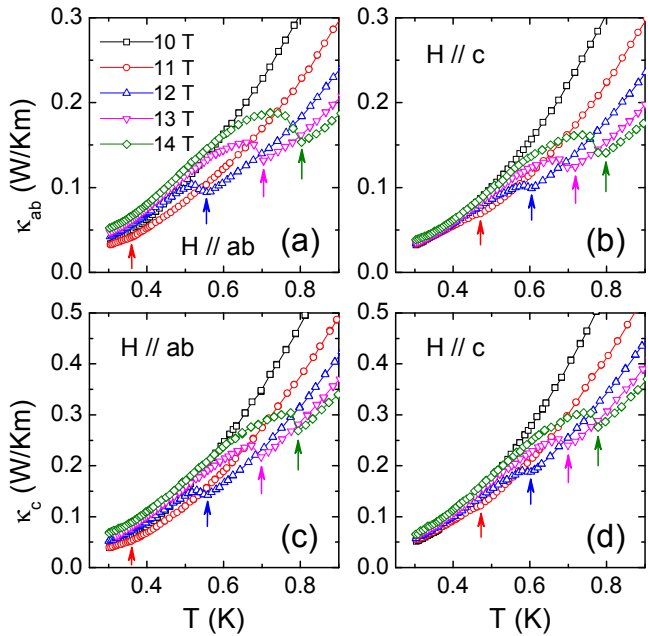}
\caption{(Color online) Temperature dependencies of the thermal
conductivity measured in magnetic field above the critical field
$H_{c1}$. The arrows indicate the critical temperatures where
$\kappa$ starts to increase.}
\end{figure}

The above are, however, not the whole story because there is only
one transition in each $\kappa(H)$ isotherm for both $H \parallel
c$ and $H \parallel ab$, whereas the former experiments revealed a
single magnetic transition and two distinct transitions,
respectively. For more precise description, we show in Fig. 5 a
series of $\kappa(T)$ curves for magnetic fields varying between
10 and 14 T. It is found that below 10 T, there is no any anomaly
in the $\kappa(T)$ curves down to 0.3 K. Above 10 T, a jump in
$\kappa(T)$ shows up and becomes bigger with increasing field up
to 14 T. From these data, we can obtain the transition points on
the $H-T$ phase diagram and compare them with the boundaries of
the magnetically ordered phases determined by heat capacity,
magnetocaloric effect, and cantilever torque
measurements,\cite{Samulon1} as shown in Fig. 6. It is interesting
that the jump-like anomalies of $\kappa$ occur at the boundary of
phase I, that is, the thermal conductivity increases at the phase
transitions to phase I from either the low-field disordered state
or the antiferromagnetically ordered phase II. This phenomenon has
a good correspondence to the specific-heat data,\cite{Samulon1}
which shows a sharp lambdalike peak and a much weaker and less
divergent peak at the phase transitions from the disordered state
or phase II to phase I and the disordered state to phase II,
respectively. It is likely that the phase transition associated
with lambdalike specific-heat anomaly has more significant
critical fluctuations that strongly scatter phonons.

\begin{figure}
\includegraphics[width=7.0cm]{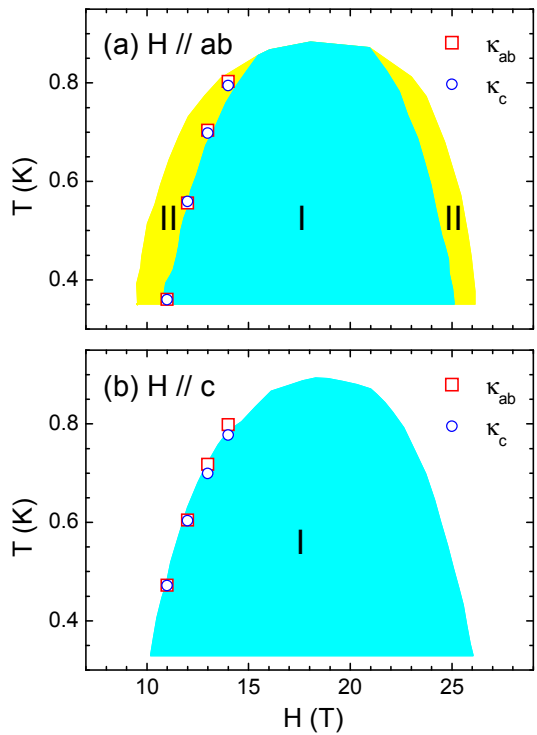}
\caption{(Color online) The transition points of $\kappa(T)$
curves from Fig. 3. Temperature-field phase diagram, obtained by
heat capacity and magnetocaloric effect measurements from Ref.
\onlinecite{Samulon1}, is shown for comparison. I and II indicate
two different phases of the triplet condensate.\cite{Samulon1} }
\end{figure}

\section{Summary}

We study the heat transport of Ba$_3$Mn$_2$O$_8$ single crystals
at very low temperatures and in high magnetic fields to probe the
field-induced magnetic phase transitions and the role of magnons
in the transport properties. In zero field, the low-$T$ thermal
conductivity shows a purely phononic transport behavior. In
applied strong field, the spin gap is gradually diminished and the
magnetic excitations are populated, which induces an enhanced
magnetic scattering on phonons and an overall strong suppression
of thermal conductivity with increasing field. Moreover, at very
low temperatures where the closure of spin gap results in a
long-range ordered AF state, the phonons are more significantly
scattered at the phase transitions. There is no evidence showing
the sizeable magnetic heat transport in this magnon BEC compound.

\begin{acknowledgements}

This work was supported by the Chinese Academy of Sciences, the
National Natural Science Foundation of China, and the National
Basic Research Program of China (Grants No. 2009CB929502 and No.
2011CBA00111).

\end{acknowledgements}

\end{document}